\documentclass[letterpaper]{article} %DO NOT CHANGE THIS
\usepackage{authblk}
\usepackage{times}
\usepackage{latexsym}

\usepackage{blindtext}
\usepackage{graphicx}  
\usepackage{xcolor}
\usepackage{varwidth}
\usepackage{amsmath,amssymb}
\usepackage{stmaryrd, bm}
\usepackage{mathtools}
\usepackage{breqn}
\usepackage{verbatim, booktabs, array} 
\usepackage{flushend}
\usepackage[utf8]{inputenc}
 
\usepackage{aaai19}  %Required
\usepackage{times}  %Required
\usepackage{helvet}  %Required
\usepackage{courier}  %Required
\usepackage{url}  %Required
\usepackage{graphicx}  %Required
\DeclareUnicodeCharacter{200E}{ }
\newcommand\citeA[1]{\citeauthor{#1} (\citeyear{#1})}

\newcommand\blfootnote[1]{%
	\begingroup
	\renewcommand\thefootnote{}\footnote{#1}%
	\addtocounter{footnote}{-1}%
	\endgroup}

\begin{document}
\title{Understanding Actors and Evaluating Personae with Gaussian Embeddings}
\author[1*]{\bf Hannah Kim}
\author[2*]{\bf Denys Katerenchuk}
\author[3]{\bf Daniel Billet}
\author[4]{\bf Jun Huan}
\author[1]{\bf Haesun Park}
\author[4*]{\bf Boyang Li}
\affil[1]{Georgia Institute of Technology \quad \textsuperscript{2}The City University of New York}
\affil[3]{Independent Actor \quad \textsuperscript{4}Big Data Lab, Baidu Research}
%s\affil[3]{}
\affil[ ]{\small \texttt {hannahkim@gatech.edu,  dkaterenchuk@gradcenter.cuny.edu}}
\affil[ ]{\small \texttt {dbillet@gmail.com, hpark@cc.gatech.edu, \{huanjun, boyangli\}@baidu.com}}

\maketitle
\begin{abstract}
Understanding narrative content has become an increasingly popular topic. Nonetheless, research on identifying common types of narrative characters, or \emph{personae}, is impeded by the lack of automatic and broad-coverage evaluation methods. We argue that computationally modeling actors provides benefits, including novel evaluation mechanisms for personae. Specifically, we propose two actor-modeling tasks, cast prediction and versatility ranking, which can capture complementary aspects of the relation between actors and the characters they portray. For an actor model, 
we present a technique for embedding actors, movies, character roles, genres, and descriptive keywords as Gaussian distributions and translation vectors, where the Gaussian variance corresponds to actors' versatility. Empirical results indicate that (1) the technique considerably outperforms TransE \cite{TransE2013} and ablation baselines and (2) automatically identified persona topics \cite{Bamman2013} yield statistically significant improvements in both tasks, whereas simplistic persona descriptors including age and gender perform inconsistently, validating prior research.
\end{abstract}

\section{Introduction}
Despite renewed interest in narrative text understanding \cite{Ouyang2015,Chaturvedi2017a}, \blfootnote{\noindent\textsuperscript{*} The main technique was conceived when Hannah Kim, Denys Katerenchuk, and Boyang Li worked at Disney Research.}the extraction of high-level narrative semantics, such as persona, remains a formidable challenge. A \emph{persona} \cite{Bamman2013,bamman2014} refers to a class of story characters that share traits, behaviors, and motivation. For example, the narratologist \citeauthor{Campbell1949} (1949) identified the hero persona, whose typical journey includes entering an unknown world, overcoming many difficulties, and receiving a reward in the end. Bamman \emph{et al}'s work (2013; 2014) pioneer computational persona discovery from narrative text. Since then, however, progress on this topic has been limited. 

A major obstacle in this line of research is the lack of automatic and broad-coverage evaluation methods. Existing work mostly relied on human evaluation, which is expensive and time-consuming to perform. Crowd workers may not possess sufficient literary understanding to provide reliable persona annotation. In comparison, literary scholars are authoritative sources of annotation but their limited availability often results in low annotation coverage. 
%Evaluation difficulties impedes further advance on persona discovery. 

Actors in movies and TV shows are an integral part of narratives, but have been largely ignored by the literature on narrative understanding. Computational modeling of actors may benefit downstream applications like movie recommendation based on actor similarity. More importantly, actors and the characters they portray are correlated. By modeling actors, we create two complimentary evaluation metrics for automatically discovered personae: \emph{cast prediction} and \emph{versatility ranking}. 

%Further, we demonstrate a representation learning technique that embeds actors, movies, and descriptive keywords as Gaussian distributions. The learned representations allow us to evaluate a given set of personae using the proposed metrics. 

%In this paper, we argue that modeling actors in narrative understanding systems can be beneficial, especially for evaluation purposes. %We can evaluate such systems using the task of \emph{cast prediction}, which asks: based on its understanding of a movie, can the system predict which actors were selected to play in the movie?
%
% In particular, cast prediction provides a test for models of story characters and automatically identified personae. 

The cast prediction task aims to predict which actors played for a particular role in a given movie. Through the controversial but ``inescapable'' \cite{Wojcik2003} practice of typecasting, many actors repeatedly play similar roles. An accurate understanding of the character roles could filter actors who never played similar characters (thereby improving precision) but keep those who did (thereby improving recall). Thus, cast prediction may be utilized as an evaluation for persona descriptors.

However, skilled actors are not limited to playing one type of characters. Versatility is defined as an actor's ability to successfully ``disappearing'' into widely divergent roles and is widely recognized as the hallmark of acting skill. 
We reckon that persona may not be effective at predicting what roles that versatile actors could play. As a remedy, we introduce a second task, versatility ranking, which aims to order actors by their versatility. Recognizing different personae played by actors should help in quantifying their versatility.  
 
As a computational model for actors and an evaluation algorithm for a given set of personae, we present a joint representation learning technique that embeds actors, movies, genres, and keywords that describe movies as Gaussian distributions \cite{Vilnis2015} and the personae as translation vectors on the Gaussian means. 
The Gaussian representation offers two advantages. First, it models the inherent uncertainty that exists in the meanings of entities and occurrence of relations in-between. For example, the meaning of user-supplied keywords is subject to users' interpretation. Casting decisions may be influenced by random events such as schedule conflicts and actor injuries. Second, there is a natural correspondence between the Gaussian variance and the actors' versatility. 

Our algorithm improves the mean rank of ground-truth answers by 332 positions over a TransE baseline \cite{TransE2013}, which  does not explicitly represent the uncertainty, underscoring the Gaussian formulation's advantage. The algorithm also achieves the accuracy of 59.72\% in versatility ranking. The difficulty of the task is noteworthy, as humans typically require professional training to accurately recognize acting skill. We experiment with a few possible persona descriptors on the two tasks. Persona topics extracted using Persona Regression \cite{Bamman2013} improve performance on both tasks, but simplistic persona descriptors including age and gender perform inconsistently, suggesting the proposed metrics can distinguish different persona descriptors.

This paper makes the following contributions:
\begin{itemize}
\item We propose an actor model that embeds movies, actors, genres, keywords, and story characters with Gaussian embeddings and translation vectors. The model outperforms baselines in cast prediction and is the first successful attempt to capture actors' versatility. 
\item We contend that cast prediction and versatility ranking can serve as novel evaluation mechanisms for automatically recognized personae. Despite significant level of noise in text preprocessing, persona topics identified from text attain statistically significant improvements on cast prediction and versatility modeling, validating prior research. 
\item We create a set of versatility rankings by four domain experts. The data and the model will be released. 
\end{itemize}

%We evaluate the embedding framework using two quantitative tasks. In the first task, we compares the computed actor versatility with judgments of professional actors. In the second task, we predict which actors play in given movies. Empirical results indicate that our framework outperforms strong baselines. 

\section{Background and Related Work}
We briefly review work on narrative understanding and unsupervised representation learning for words and concepts.
\subsection{Narratives and Story Characters}
As a complex artifact, a narrative can be analyzed and computationally modeled as multiple layers that are dependent on each other, including the surface text \cite{Reagan2016}, event semantics, and script-like patterns \cite{Li2013,Pichotta:acl16,Ferraro2016,Hu-Walker2017,Martin2018}, macro-structures \cite{Ouyang2015,Finlayson2016,Li2018}, story characters and their relations \cite{Bamman2013,bamman2014,Valls-Vargas2015,Chaturvedi2017a} as well as narrative quality \cite{Wang2017:quality,Sagarkar2018}. 

%Literary theorists and narratologists observed similarities between different characters in different narratives. 
%
%, such as Campbell's hero \cite{Campbell1949}. Archetypes are types of characters that share common motivation, traits and behaviors and frequently reappear in different scenarios under disguise. For instance, the derelict hero archetype in \cite{Schmidt2001} includes Joey Tribbianni (Matt LeBlanc) in \textit{Friends}, Joel Goodson (Tom Cruise) in \textit{Risky Business}, and Jay (Will Smith) in \textit{Men in Black}. Depending on the theory, the exact number of existing archetypes vary from Carl Jung's 12 and Schmidt's 45 \cite{Schmidt2001} to the 99 listed on tvtropes.org\footnote{\url{http://tvtropes.org/pmwiki/pmwiki.php/Main/ArchetypalCharacter}}.

%\sloppypar{
Among various facets of a narrative, many novelists consider story characters to be the most crucial for a good story \cite{Woolf1924}. A common class of characters (sometimes referred to as a persona) share motivation, traits and behaviors and are often recast under different disguise. A classic example is Campbell's hero, who receives a call for adventure, takes on many challenges before a final, most difficult challenge, and gains a boon for completing the quest \cite{Campbell1949}. Many story protagonists, especially in the adventure genre, follow this pattern. We are not aware of a definitive list of personae; theories range from Carl Jung's twelve and Schmidt's forty-five (\citeyear{Schmidt2001}) to the ninety-nine listed on tvtropes.org\footnote{\url{http://tvtropes.org/pmwiki/pmwiki.php/Main/ArchetypalCharacter}}
%}

Computational techniques that model story characters are the most relevant to this work. 
Bamman \emph{et al.} (\citeyear{Bamman2013,bamman2014}) extract persona as a group of topics, which are multinomial distributions over action verbs, possessives, and modifiers respectively. The rationale is that similar characters would perform / receive similar actions and have similar descriptions. For example, a hero often rescues, a sidekick would banter, and a villain tends to be angry. 
\citeA{Bamman2013} adopt a logistic prior uses the actor's age, gender, and movie genre to aid the assignment of movie characters to personae. \citeA{bamman2014} further incorporate the influence of the author and the background topic frequency into a log-linear topic model \cite{Eisenstein2011}. 
\citeA{Valls-Vargas2015} propose a system architecture with feedback loops for identifying personae from Russian folklore. \citeA{Chaturvedi2017a} employ hidden Markov models to learn how relationships evolve between characters. 

This paper does not attempt to recognize personae directly. Instead, we take a list of recognized personae as input. Since many actors play similar characters in their career, a good list of personae should facilitate the modeling of actors. The two tasks we propose can serve as evaluation metrics for personae recognized unsupervisedly. 

Different evaluation methods have been employed for story characters and their relations. \citeA{Bamman2013} use gold clusters containing the same character in sequels or remakes and crowd-supplied examples for TV tropes.  \citeA{bamman2014} employ 29 hypotheses created by a literary scholar regarding character similarity. \citeA{Chaturvedi2017a} use about 400 sentences from story summaries as their test for binary relation classification, as well as a relation analogy task. The cast prediction task is similar to the same-character clusters in its objectivity but arguably has better coverage as it is not limited to movies that share characters. 

%In the economics and marketing literature, many statistical analysis have been performed on identifying causal factors contributing to box-office success and award wins. For example, \cite{Pardoe2008} predicted Oscar winners using a Bayesian logic model. De Vany and Walls \cite{DeVany1999} found box-office revenue to have a Pareto distribution with infinite variance, suggesting revenue forecasting is difficult. This topic is out of the scope of the current paper, so we refer interested readers to a survey \cite{Kumb2007}. It is worth noting that such studies tend to rely on relatively simplistic models for actors, often a dichotomy of stars and non-stars, and ignore similarities between actors. This suggests a more sophisticated model of actors is likely to have broad impact beyond machine learning and computer science.  

\subsection{Unsupervised Lexical and Conceptual Representations}
Distributed word representations are usually learned based on word co-occurrence statistics \cite{Mikolov2013,Pennington2014,Ling2015}. \citeauthor{Vilnis2015} (\citeyear{Vilnis2015}) generalize word vectors to Gaussian distributions that occupy a volume in space. \citeA{Athiwaratkun2017} propose mixtures of Gaussian embeddings to account for multiple word senses. 

Despite broad adoption, word embeddings acquired from text do not capture all types of semantic information. \citeauthor{Rubinstein2015} (\citeyear{Rubinstein2015}) find them particularly deficient in taxonomic and attributive information (e.g., a lemon is a fruit and typically green or yellow). 
In comparison, such information is well captured by ontologies and knowledge graphs, where a relation $r$ between entities $h$ and $t$ is represented as $\langle h, r, t \rangle$. For example, one fact above can be represented as the tuple $\langle$\emph{lemon}, IsA, \emph{fruit}$\rangle$.

Various techniques \cite{TransE2013,He2015,Nickel2015} learn embeddings from ontologies and knowledge graphs. The embeddings are evaluated by filling in missing items in relation tuples such as $\langle$\emph{lemon}, IsA, \underline{\hspace{0.6cm}}$\rangle$. 
In particular, the TransE formulation~\cite{TransE2013} assigns embeddings $v_h$, $v_r$, and $v_t$ respectively to the entities $h, t$ and the relation $r$. For every ground-truth tuple $\langle h, r, t \rangle$, the optimization attempts to achieve $v_h + v_r = v_t$. \citeA{He2015} apply Gaussian embeddings to knowledge graphs and represent both entities and relations as Gaussian distributions. In this work, relations are represented as translation vectors on the Gaussian means and are not associated with covariance matrices. 

%The entities and relations considered in this paper can be seen as a knowledge graph. 
Different from existing work that take the knowledge graph as given, we investigate what kind of role descriptors that can best describe the interaction between actors and movies, as measured by cast prediction accuracy. 
Moreover, we identify actor versatility as a natural grounding for the learned variance of actors, whereas in most previous work the learned embeddings lack practical interpretation.

\section{Learning Joint Gaussian Embeddings}

We aim to build a model that connects movies with actors and the types of roles that they played. To help anchor the movies relative to each other, we also consider descriptors of movies including genres and keywords that can be acquired from many websites or crowd annotation. 

More formally, the data contain the set of movie-persona-actor triples $\mathcal{D}_{mpa} = \{\langle m_i, p_i, a_i \rangle\}$ and the set of movie-keyword pairs $\mathcal{D}_{mk} = \{\langle m_i, k_i\rangle\}$. 

%There are usually 20 or so wide recognized genres (e.g., \emph{action}, \emph{comedy}). In addition, websites often allow the users to define their own keywords. Examples  (e.g., \emph{coming-of-age},  \emph{rescue-mission}) and other descriptors (e.g., \emph{tragic-hero},  \emph{critically-bashed}). 

%\sloppypar{
Genres and keywords may have varying degrees of specificity. For instance, the genre \emph{drama} is generic, whereas the keyword \emph{younger-women-older-man-relationship} is more specific. 
Netflix is known to use a large genre hierarchy\footnote{\url{http://ogres-crypt.com/public/NetFlix-Streaming-Genres2.html}} ranging from broad genres to highly specific ones such as \emph{African-American-showbiz-dramas}. 
Thus, explicitly representing the uncertainty of a keyword's meaning could be advantageous. For simplicity, we henceforth do not distinguish between genres and keywords and call both keywords. 
 For persona information, we adopt the persona topics identified by Persona Regression \cite{Bamman2013} and the age / gender of the actors at the time of movie release. 
%}

\begin{table}[t]
	\centering
	\caption{Key mathematical notations used in this paper. }
	\renewcommand{\arraystretch}{1.2}
	\label{tab:notation}
	\begin{tabular}{lp{5.2cm}}
		\toprule
		$D$ & The embedding space's dimension \\
		$m_i, \bm \mu^m_i, \Sigma^m_i$ & The $i^{\text{th}}$ movie, its mean vector $\bm \mu^m_i$ and covariance matrix $\Sigma^m_i$ \\
		$a_j,  \bm \mu^a_j, \Sigma^a_j$ & The $j^{\text{th}}$ actor, its mean vector $\bm \mu^a_i$ and covariance matrix $\Sigma^a_i$  \\
		$k_r, \bm \mu^k_r, \Sigma^k_r$ & The $r^{\text{th}}$ keyword, its mean vector, and covariance matrix. \\
		$p_s, \bm \nu^p_s$ & The $s^{\text{th}}$ persona, represented by a translation vector $\bm \nu^p_s$ \\
		$\langle m_i, p_i, a_i \rangle$ & A real movie-persona-actor  triple, \emph{i.e.}, the actor $a_i$ played the role of $p_i$ in the movie $m_i$.\\
		$\langle m_i, p_i, a^{-}\rangle$ & A negative triple, \emph{i.e.}, the actor $a^{-}$ did \textbf{not} play the role of $p_i$ in movie $m_i$.\\
		$\langle m_i, k_i \rangle$ & A real movie-keyword  pair, \emph{i.e.}, the movie $m_i$ is labeled with the keyword $k_i$.\\
		$\langle m_i, k^{-} \rangle$ & A negative pair, \emph{i.e.}, the movie $m_i$ is \textbf{not} labeled with the keyword $k^-$.\\
		$\mathcal{N}(\bm x; \bm \mu, \Sigma)$ & Gaussian pdf for $\bm x$ under the mean $\bm \mu$ and the covariance $\Sigma$. \\
		$diag(\bm v)$ & A diagonal matrix with $\bm v$ on its diagonal and zero everywhere else. \\
		\bottomrule
	\end{tabular}
\end{table}

\sloppypar{
The central feature of our model is that every movie, actor, and keyword (but not persona) is represented as a Gaussian distribution instead of a single vector. The Gaussians are in $D$-dimensional space, with mean $\bm \mu$ and covariance matrix $\Sigma$. We use superscript for further differentiation: $\bm \mu_i^m \in \mathbb{R}^D$ and $\Sigma_i^m \in \mathbb{R}^{D\times D}$ denote the parameters for movies. $\bm \mu_j^a \in \mathbb{R}^D$ and $\Sigma_j^a \in \mathbb{R}^{D\times D}$ denote those for actors, and $\bm \mu_r^k \in \mathbb{R}^D$ and $\Sigma_r^k \in \mathbb{R}^{D\times D}$ denote those for keywords. Every persona, in contrast, is represented as a single vector $\bm \nu_s^p$. See Table \ref{tab:notation} for a summary of key notations.}

For simplicity, we use only the diagonal entries of the covariance matrices. We assign a vector $\bm \sigma^2_i = (\sigma_{i1}^2, \sigma_{i2}^2, \ldots, \sigma_{iD}^2)$ to each entity and let its covariance matrix $\Sigma_i = diag(\bm \sigma_i^2)$.  Preliminary experiments indicate a spherical setup, where all variances $\sigma_{id}$ are the same, yields the best performance. 

\subsection{Model}

Following \citeA{Vilnis2015}, we define the symmetric similarity between a movie $m_i$ and a keyword $k_r$ as an integral over the product of two Gaussian pdfs:
\begin{equation} \label{eq-movie-keyword-sim}
\begin{split}
S(m_i, k_r)  & =\log \int \mathcal{N}(\bm x; \bm \mu^m_i, \Sigma^m_i) \mathcal{N}(\bm x; \bm \mu^k_r, \bm \Sigma^k_r) \,d{\bm x}  \\
 & = \log \mathcal{N}(\bm \mu^k_r; \bm \mu^m_i, \Sigma^m_i + \Sigma^k_r) \\
 & = \log \mathcal{N}(0; \bm \mu^m_i - \bm \mu^k_r, \Sigma^m_i + \Sigma^k_r) \\
 & = -\frac{1}{2} \log \text{det}(\Sigma^m_i + \Sigma^k_r) + \\
& -\frac{1}{2} (\bm \mu^m_i - \bm \mu^k_r)^\top (\Sigma^m_i + \Sigma^k_r)^{-1} (\bm \mu^m_i - \bm \mu^k_r) + c
\end{split}
\end{equation}
where $c$ is a constant that does not matter for the optimization. The similarity function has an intuitive geometric interpretation: to maximize the similarity, we can reduce the distance between the two means, or increase the diagonal variance to make the Gaussian less concentrated. At the same time, the determinant term works like a regularizer that prevents the diagonal variances from exploding. For the diagonal covariance matrix $\Sigma$, $-\log \text{det}\Sigma = -\log \prod_{d=1}^D \sigma^2_{d}$. Maximizing this term pushes the individual variance terms $\sigma^2_{d}$ closer to zero.

\begin{figure}[t]
  \includegraphics[width=\linewidth]{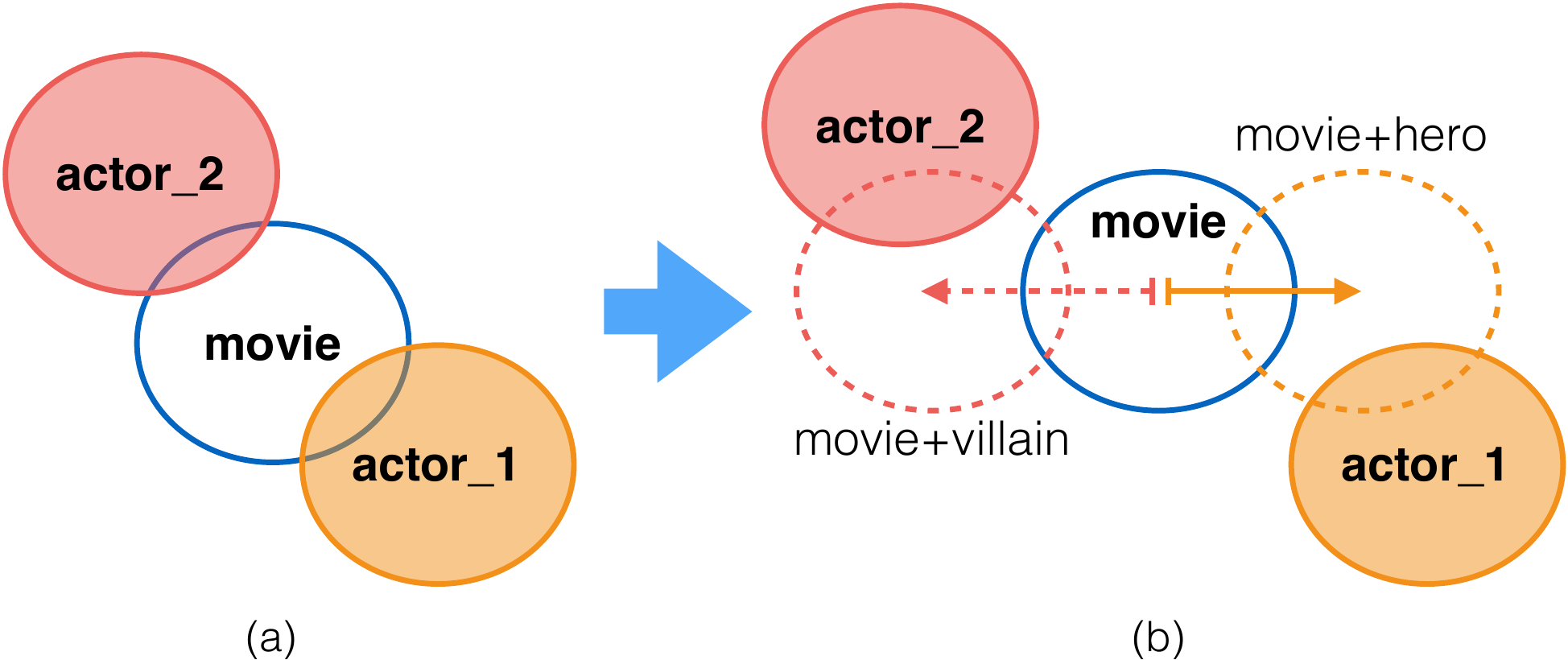}
  \caption{The intuition underlying the persona translation vectors. The shaded circles represent Gaussian distributions for two actors playing the hero and the villain respectively in a movie. The dashed circles represent the movie Gaussian after the translation vectors in opposite directions are applied to its mean. Comparing (b) with (a), the actors are further separated after the translation vectors are applied.}
  \label{fig:role-vector intuition}
\end{figure}

The similarity between an actor and a movie has two possible definitions, depending on if we incorporate persona information. Similar to Eq. \ref{eq-movie-keyword-sim}, we can define a persona-free similarity between movie $m_i$ and actor $a_j$ as
\begin{equation}
\label{eq:pf}
S_{\text{pf}}(m_i, a_j) = \log N(\bm \mu^m_i; \bm \mu^a_j, \Sigma^m_i + \Sigma^a_j)
\end{equation}
In the second definition, we further introduce a translation vector $\bm \nu^p_s \in \mathbb{R}^D$ for every persona $p_s$. The similarity for movie $m_i$, persona $p_s$, and actor $a_j$ becomes
\begin{equation} 
\label{eq:psim}
S_{\text{p}}(m_i, p_s, a_j) = \log N(\bm \mu^a_j; \bm \mu^m_i + \bm \nu^p_s, \Sigma^a_j + \Sigma^m_i)
\end{equation}
Figure \ref{fig:role-vector intuition} illustrates this intuition behind this use of translation vectors: actors playing different character roles in the same movie should be separated from each other while remaining close to the movie. For example, a hero in a science fiction movie should be separated from a villain from the same movie, but they should stay in the proximity of the movie, which in turn is close to the science fiction genre. 

For the complex concept of persona, we allow multiple descriptors, each with one vector. We denote the $z^{\text{th}}$ descriptor of the $s^{\text{th}}$ persona as $\bm \nu_{s, z}$. The final persona vector $\bm \nu^p_s$ is computed as the sum $\sum_z \bm \nu_{s, z}$ or the concatenation $[\bm \nu_{s, 1}, \ldots, \bm \nu_{s, Z}]$. In experiments, we adopt the sum of three descriptors: age, gender, and persona topics.

\subsection{Loss Functions}
For movies and keywords, we define the margin loss $\mathcal{L}_{mk}$ as,
\begin{equation}
\label{eq_loss_mk}
\mathcal{L}_{mk} = \sum_{\langle m_i,k_i \rangle \in \mathcal{D}_{mk}} \mathop{{}\mathbb{E}}_{k^{-}\neq k_i}
\llbracket 
g(S(m_i, k_i), S(m_i, k^{-})) 
\rrbracket 
\end{equation}
\begin{equation}
g(s_1, s_2) = \max(0, \phi - s_1 + s_2)
\end{equation}
where $\phi$ denotes the margin of separation and $k^{-}$ is a randomly sampled  negative keyword that does not co-occur with $m_i$. 
For a real pair $\langle m_i,k_i \rangle \in \mathcal{D}_{mk}$ and a negative pair $\langle m_i,k^{-} \rangle$, the loss is minimized when $S(m_i, k_i) - S(m_i, k^{-}) \geq \phi$, so that the similarities of real and negative data are separated by the margin. The expectation over $k^{-}$ is approximated by random sampling during every epoch.

Similarly, for the real movie-actor-persona triple $\langle m_i, p_i, a_i \rangle$ and a negatively sampled actor $a^{-}$ , we define the loss $\mathcal{L}_{mpa}$ as,
\begin{equation}
\label{eq_loss_mpa}
\mathcal{L}_{mpa} = \sum_{\langle m_i, p_i, a_i \rangle \in \mathcal{D}_{mpa}} \mathop{{}\mathbb{E}}_{a^{-}\neq a_i}\llbracket g(S(m_i, p_i, a_i), \\S(m_i, p_i, a^{-}) )\rrbracket
\end{equation}
The final loss function to be minimized is the sum $\mathcal{L} = \mathcal{L}_{mk} + \mathcal{L}_{mpa}$. Intuitively, this optimization pushes co-occurring entity pairs (or triplets) toward each other and pushes pairs (or triplets) that do not co-occur further apart, so that the learned similarity function reflects the observed co-occurrence data. 
% It is possible to weigh $\mathcal{L}_{mk}$ and $\mathcal{L}_{mpa}$ differently, but here we use equal weights.
%That is, movies that share keywords are pushed together and those that do not share keywords are pushed apart; actors that play similar roles in similar movies are also pushed together. 

\subsection{Regularization and Learning Rate Decay}
Compared to the enormous corpora for training word embeddings, the data for movies and actors are drastically more scarce. To bridge the gap, we devise a form of dropout regularization \cite{Hinton2012:Dropout}. For each pair or triplet of entities in the loss function (Eq. \ref{eq_loss_mk} and Eq. \ref{eq_loss_mpa}), we create a $D$-dimensional random vector $\bm q$, whose components $q_i \in \{0, 1/p^0\}$ are independently drawn as 0-1 Bernoulli trials with probability $p^0$. $\bm q$ is component-wise multiplied to all mean vectors during training and redrawn every epoch. This is similar to ignoring some dimensions when computing the similarity for an individual data point.  

To avoid numerical issues introduced by the inverse matrix and stabilize training, we introduce an upper bound $\sigma^2_{max}$ and lower bound $\sigma^2_{min}$ for the diagonal variances. Values that exceed the given range are clipped. 

Moreover, we adopt the cosine schedule for learning rate decay \cite{Loshchilov2017}. For the $t^{\text{th}}$ mini-batch, the learning rate $\eta_t$ is given by
\begin{equation}
\eta_t = \eta_{min} +  \frac{1}{2} \left( \cos \left(\frac{t}{T_{max}}\right) + 1 \right) \left(\eta_0 - \eta_{min} \right)
\end{equation}
where $\eta_0$ is the initial learning rate,  $\eta_{min}$ the minimum learning rate, and $T_{max}$ the total number of mini-batches used during training. The cosine schedule achieves slow decay at the beginning and the end of training but fast decay in the middle, in comparison to the linear schedule.
%\begin{equation}
%\eta_t  = \eta_{min} + \left(\eta_0 - \eta_{min} \right) \left(1 - \frac{t}{T_{max}} \right)
%\end{equation}
Early experiments show the cosine decay improves training when the range between $\sigma^2_{max}$ and $\sigma^2_{min}$ is large (\emph{e.g.}, 100 to 0.0001) and helps in differentiating actors with diverse versatility.

%\subsection{Identifying Character Archetypes}
%A typical assumption in identifying character archetypes is that characters can be defined by their behaviors and how they are described in the text \cite{Bamman2013}. Following this intuition, we first perform coherence resolution to link pronouns to the story character. After that, we identify actions that are performed by the character, actions that are performed to the character, as well as descriptions of the characters. The actions are represented by the verbs and the descriptions are summarized as nouns and adjectives. These operations result in a list of words in each category. 

%Recently, distributed word representations like word2vec and GLoVe have been shown to capture lexical meaning of words. Here we adopt counter-fitted word embeddings \cite{Mrksic:2016} that have been shown to highly correlate with human judgments of word similarity. For each category, we obtain an average of the word embeddings. Their concatenation becomes the representation of the literary character. As a final step, we perform clustering on the character representations to identify archetypes. 

\section{Experiments}
\label{sec:experiments}
The experiments include two tasks. The supervised task of cast prediction investigates if the model can predict which actors played a given role in a given movie. In unsupervised versatility ranking, we examine the degree to which the learned variances of the actor Gaussians agree with the versatility judgments from human experts. In the ablation study, we compare the performance of different persona descriptors. 

%In the Results sections below, we describe task-specific baselines and the best performing set of hyperparameters. 

\subsection{Data and Preprocessing}
\label{sec:experiment-setup}
With permission from The Movie Database\footnote{\url{www.themoviedb.org}}, we collected the metadata of 335,037 movies with complete cast lists, genres, and user-supplied keywords. 
We kept prominent actors that appear in the first four leading roles of the cast and entities that appear in at least 10 relations. This procedure yields 2,200 actors in 15,997 movies, 2,213 user-supplied keywords, and 20 curated genres.

To collect persona information, we perform Persona Regression \cite{Bamman2013} on Wikipedia plot summaries that match the collected movies on metadata including title, year, and IMDB identifiers. Every story character is assigned to one of 50 latent persona topic groups. Due to sparsity in the plot summaries and imperfect coreference resolution, we are only able to assign a topic group to 30.5\% of the 38,179 movie-actor pairs. 

The age and gender of actors can also be considered as persona descriptors. For example, the protagonist often receives advice from someone more mature (\emph{e.g.}, the Sage in Carl Jung's 12 characters). Persona Regression makes use of age and gender in the prior. For this paper, we extract the gender and the age of actors at the time of the movie release from TMDB and Freebase~\cite{Bollacker:2008}. The range of age (0-120) is discretized into 5-year segments.

We compute the persona vector $\nu_s$ as the sum of three vectors $\nu_{s, topic}$, $\nu_{s, age}$, $\nu_{s, gender} \in \mathbb{R}^D$, which respectively represent the learned persona topic group (T) for the character in the plot summary, the actor's age (A), and the actor's gender (G). Every unique value in the three categories is mapped to a vector learned during training. 

\subsection{Ablated Baselines}

We create several ablated versions of the proposed model. In the first baseline, which we name Joint Gaussian Embeddings (JGE), we ignore persona information. The second (JGE+T) supplement JGE with the persona topic groups identified by Persona Regression. The third (JGE+AG) employs only age and gender information as the persona, but ignore the topic groups. The full joint Gaussian embedding model, which uses all persona information, is denoted as JGE+AGT. For each version, we also try initializing the mean of keyword Gaussians with GloVe word embeddings~\cite{Pennington2014}. For multiword expressions, we sum the individual word embeddings. The JGE baseline uses the persona-free similarity function in Eq. \ref{eq:pf}, but others use the similarity defined in Eq. \ref{eq:psim}. 

\begin{table}[t]
\centering
\caption{Mean rank and Hits@10 scores for the link prediction task. The third column shows improvements on mean rank with respect to different models indexed by line numbers in brackets. * denotes statistically significant improvements of the best model JGE+AGT over JGE+AG.}
\renewcommand{\arraystretch}{1.2}
\vspace{0.1in}
\label{tab:link-prediction}
\begin{tabular}{llll}
\toprule
               & Mean Rank &  Hits@10 & MR Change\\ 
\midrule
(1) TransE         & 506.75    &  3.73 \%   \\ 
\midrule
(2) JGE            & 479.59	& 6.17 \%   & -27.2 w.r.t (1)\\ 
(3) ...+GloVe      & 478.48    &  6.12 \%  & -28.3 w.r.t (1)\\ 
(4) JGE+T          & 479.66    &  6.23 \%  & 0.07 w.r.t (2)\\ 
(5) ...+GloVe    & 471.99   & 6.13 \%  &  -1.13 w.r.t (3)\\ 
(6) JGE+AG         & 176.66	& 12.52 \%  &  -303 w.r.t (4)\\ 
(7) ...+GloVe   & 176.77	& \textbf{12.69} \%  & -295 w.r.t (5)\\ 
(8) JGE+AGT        & \textbf{174.64}	& 12.54 \%  & -2.0 w.r.t (6)*\\ 
(9) ...+GloVe  & 175.48	& 12.63 \%  & -1.3 w.r.t (7)\\ 
\bottomrule
\end{tabular}
\end{table}

\subsection{Cast Prediction: Setup and Results}
The task-specific baseline we adopt is the popular TransE graph embedding algorithm~\cite{TransE2013}. The movie-persona-actor triples are randomly split into a 70\% training set, a 15\% validation set, and a 15\% test set. All movie-keyword pairs are in the training set. The hyperparameters of all methods, including training epochs, embedding dimension, learning rate, margin, mini-batch size and so on, are extensively tuned on the validation set. After validation, we choose the best set of hyperparameters, train a new model using both training and validation data, and report the performance on test data. 

The selected hyperparameters are as follows. The dimension $D$ is set to 40, margin $\phi$ is 4, batch size is 128, and dropout probability is 0.6. The learning rate starts at 0.15 and is reduced to 0.0001 over 600 epochs. The optimization is performed using RMSProp. For the AG version, $\sigma_{min}$ is set to 0.0001 and $\sigma_{max}$ is set to 25. For every other method, $\sigma_{max}$ is set to 100. We report performance average of 20 runs with random initialization.

Table \ref{tab:link-prediction} shows two performance measures: the mean rank and hits@10. Mean rank is defined as the average position of all ground-truth answers in the ranked list of all possible answers. Hits@10 is defined as the percentage of correct answers among the top 10 highest ranked answers. That is, mean rank evaluates the entire ranked list, whereas hits@10 focuses only on the highest ranked portion, so we consider mean rank to be a more holistic measure than hits@10.

Our best model outperforms TransE by 332.11 on mean rank and 8.81\% on hits@10. Even the basic JGE model achieves an improvement of 27.16 on mean rank and 2.44\% on hits@10 over TransE. 
Among the ablated versions, JGE+AGT attains the best mean rank of 174.64 and JGE+AG+GloVe the best hits@10 of 12.69\%. 
The difference between the best model, JGE+AGT, and the corresponding second best, JGE+AG, is statistically significant with $p<0.05$ under one-tailed Welch's t-test.

\subsection{Cast Prediction: Discussion}

Our method achieves significant performance improvement over the TransE baseline, which makes use of translation vectors but does not represent the uncertainty in the concepts. This suggests the Gaussian formulation is advantageous for modeling the relations between movies, actors, and keywords.  

The benefits of persona descriptors deserve detailed analysis. Using age and gender information roughly doubles prediction accuracy over the JGE models.  The finding is consistent with the general observation that actor choices are limited by the roles they need to portray. 

Additionally, under severe data sparsity (only 30.5\% movie-actor pairs have a persona topic) and substantial errors in text parsing the and coreference resolution, the identified persona topics still improve mean rank, which is a more holistic measure than hits@10. 
It is also worth noting that we utilize far less information than the actual casting process, which would consider complete actor profiles including weight, height, accents and so forth as well as auditions and the actors' schedules. We find the algorithm's performance under existing data encouraging. 
To sum up, the results show that (1) the model successfully captures some aspects of the cast selection process and (2) good understanding of character persona can indeed improve the performance of cast prediction, corroborating our claim that cast prediction can serve as an evaluation metric for automatic persona discovery.

\subsection{Actor Versatility: Data Collection and Setup}
In the second evaluation task, we study if the learned Gaussian variance of actors corresponds to actor's versatility ranked by human experts. We posit that actors with wider acting spectra would have larger Gaussian variance than typecast actors, e.g., actors who play only action heroes.

We collected versatility rankings from four judges who received formal education in acting and had more than 5 years of acting experiences in Hollywood. None of them was aware of our model's output. 
We took two measures to reduce ambiguity in the judges' answers. First, in order to make sure the judges are familiar with the actors, we selected 250 actors that are most searched for on Google\footnote{\url{www.google.com/trends/}} and were likely well-known. Second, we used relative ranking instead of Likert scale. 

We created a list containing 200 sets of actors, where 5 actors (among 250) were randomly selected for a set (i.e., an actor can belong to multiple sets).
Every judge received the same list and was asked to rank actors in each set in versatility. They were allowed to skip any sets or give identical ranks to actors.
We break down each set to ${5 \choose 2}=10$ pairwise comparisons. From 200 sets, there is a total of 987 unique pairs. The judges appeared to be very consistent with themselves; three were inconsistent on only one pair of actors and one was consistent on all pairs. There is unanimous agreement on 534 or 54.1\% of pairs. On 318 pairs, or 32.2\%, one judge differed with the rest. Fleiss' Kappa is 0.498, indicating fair agreement. Given the vagueness and subjective nature of the question, we believe the agreement is decent. For evaluation, we only use 861 pairs where a majority decision is reached. 

In creating the validation-test split, a potential issue is that test data may sneak into the validation set due to the transitivity in the rankings. Consider the triplet relations $a\succ b\succ c$, if $a\succ b$ and $b\succ c$ both appear in the validation set, and $a\succ c$ appears in the test set, we would have seen test data during validation.

We prevent the issue above by distributing the actors evenly into two disjoint sets $\mathcal{A}_{val}$ and $\mathcal{A}_{test}$. Pairwise rankings between actors in $\mathcal{A}_{val}$ are used as the validation set. All other rankings are used as the test set. That is, the test set is a subset of $\mathcal{A}_{val} \times \mathcal{A}_{test} \cup \mathcal{A}_{test} \times \mathcal{A}_{test}$. As a result, for any ranking $a\succ b$ in the test set, at least one actor does not appear in the validation set.
In training, we use all available relation data, but not the versatility rankings. Thus, this task is unsupervised.

We created three heuristic baselines for this task. For the first baseline (Genre), we compute the frequency that an actor appears in every genre. The entropy of the count vector is used as an indication for versatility. An actor with higher entropy is considered more versatile. 
We repeat the entropy computation for the automatically identified persona topic groups (PTG) as well as topics of keywords (Keyword-Topics), where topics are found by non-negative matrix factorization~\cite{kim11nmf}.

\begin{figure*}[t]
  \includegraphics[width=\linewidth]{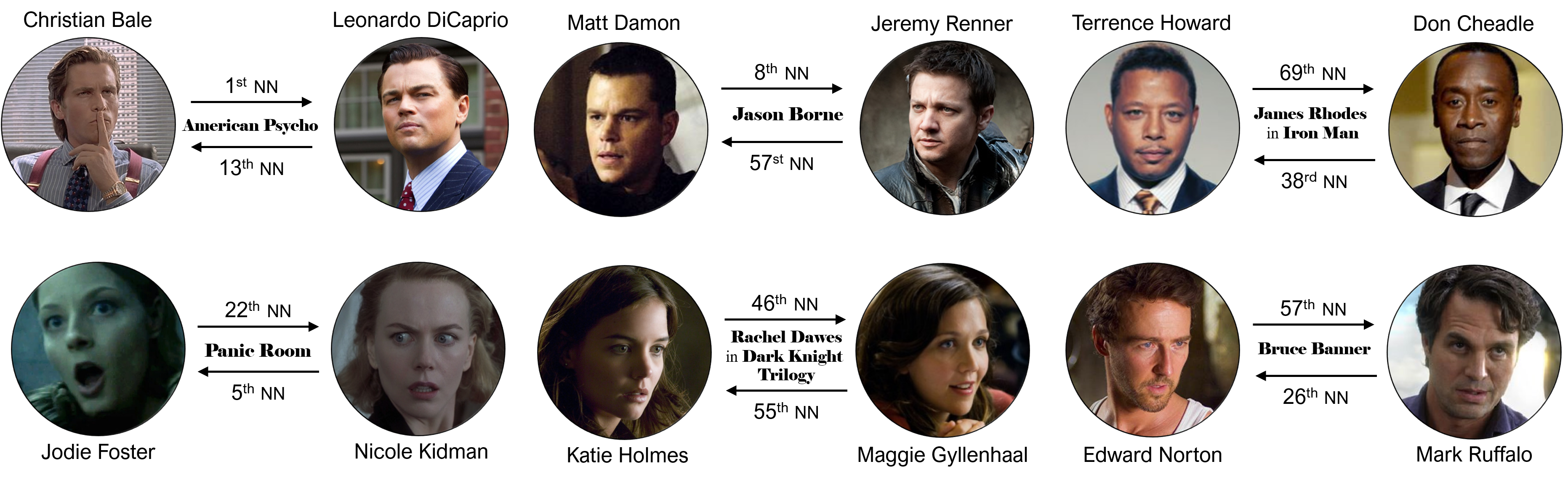}
  \caption{Movies and movie series that replace one actor with another hint at the similarity between the two. We show the nearest neighbor ranking between the two actors, computed by our best performing model for cast prediction.}
  \label{fig:qualeval}
\end{figure*}

%\texttt{\begin{table*}[t]
%	\centering
%	\caption{Performance of ranking actors' versatility on the validation set and the test set. }
%	\renewcommand{\arraystretch}{1.2}
%	\vspace{0.1in}
%	\label{tab:v-valid}
%	\begin{tabular}{lllll}
%		\toprule
%		& \multicolumn{2}{c}{Validation} & \multicolumn{2}{c}{Test} \\ \cmidrule(lr){2-3} \cmidrule(lr){4-5}
%		& Accuracy (\%) & Rank Corr. & Accuracy (\%) & Rank Corr. \\ 
%		\midrule
%		Genre          & 45.68 & -0.063  & 43.66 & -0.118 \\ 
%		Keyword-Topics  & 37.65 & -0.148 & 39.82 & -0.154  \\
%		PTG            & 50.00 &  -0.145  & 33.33 & -0.091 \\ \midrule
%		JGE            & 55.13 &	0.0697 & 55.60 & 0.085 \\ 
%		... +Glove      & 45.68 & 0.032 &  55.35 &	0.068 \\
%		JGE+T          &  \textbf{62.30} & \textbf{0.201} & \textbf{60.03} & \textbf{0.165}  \\ 
%		... +Glove    & 62.16 &	0.192	& 59.31 &	0.162      \\ 
%		JGE+AG         & 56.95 &	0.096 &	51.07 &	0.059    \\ 
%		... +Glove   & 58.97 & 0.135 & 53.14 & 0.079   \\ 
%		JGE+AGT        & 55.82 &	0.135 &	51.23 &	0.054    \\ 
%		... +Glove  & 56.57 &	0.093 & 50.57 & 0.033 \\ 
%		\bottomrule
%	\end{tabular}
%\end{table*}}

\begin{table}[t]
	\centering
	\caption{Performance of ranking actors' versatility, including pairwise accuracy on the validation and test sets and rank correlation on the test set. * denotes statistically significant differences between the best models and their corresponding second-best baselines. }
	\renewcommand{\arraystretch}{1.2}
	\vspace{0.1in}
	\label{tab:v-valid}
	\begin{tabular}{lp{1.4cm}p{1.4cm}p{1.4cm}}
		\toprule
		& Val. Acc. (\%) & Test Acc. (\%) & Test Rank Corr. \\ 
		\midrule
		Genre          & 42.72 & 45.60 & -0.082  \\ 
		Keyword-Topics & 34.74  & 39.78 & -0.192  \\
		PTG            & 43.48 & 43.14 &  -0.196 \\ \midrule
		JGE            & 47.77 & 55.13 &	0.070  \\ 
		... +Glove      & 46.71 & 55.06 & 0.072 \\
		JGE+T          & 61.78 & \textbf{59.72}* & 0.163* \\ 
		... +Glove    & 62.30 & 59.39* &	\textbf{0.165}*  \\ 
		JGE+AG        & 56.95 & 50.84 &	0.059 \\ 
		... +Glove   & 58.05 & 52.95 & 0.084   \\ 
		JGE+AGT        & 56.22 & 50.57 &	0.043    \\ 
		... +Glove  & 55.33 & 50.53 &	0.039 \\ 
		\bottomrule
	\end{tabular}
\end{table}

\subsection{Actor Versatility: Results and Discussion}

Table \ref{tab:v-valid} shows prediction accuracy for pairwise decisions and rank correlation averaged over 20 runs. Since the expert rankings provide only a partial order, we generate 100 topological sorts with random tie breaking and compute the average rank correlation. The two measures do not perfectly align because (1) random tie-breaking introduced some fluctuations and (2) pairwise comparisons do not carry the same weight in rank correlation as some pairwise rankings imply others by transitivity. 
The hyperparameters for the best performing model (JGE+T) are the same as the previous task except that the initial learning rate is set to 0.01, $\sigma_{min}$ is 0.0001, and $\sigma_{max}$ is 50.

The three heuristic baselines fall below random chance, indicating the problem is non-trivial. The best versatility prediction is obtained by the JGE+T model with only automatically extracted persona topics. JGE is the second best. Under one-tailed Welch's t-test, the difference between JGE+T and JGE, as well as that between JGE+T+GloVe and JGE+GloVe, are statistically significant with $p<10^{-5}$. 

Interestingly, while persona topics remain beneficial,
the use of age and gender information hurts performance, reversing the observation from cast prediction. It is likely that age and gender do not correlate well with an actor's skill, even though they could practically limit the range of roles the actor can play. Worse still, the additional parameters for age and gender vectors contribute to more overfitting to the validation set. 

Overall, we attain significant improvements over baselines in an unsupervised manner with far less information than to which the human judges have access. Automatically identified persona topics boost performance by 4.59\% over the second best ablation and 6.77\% over the best baseline using age and gender. Together with the cast prediction task, the results demonstrate that automatically recognized persona topics can enhance the modeling of actors and movies. The different performance of age and gender suggests that the two tasks reveal different aspects of persona and are likely complementary evaluation metrics.

\subsection{Qualitative Evaluation: Actor Replacement}
Substitution between actors provides a rare opportunity to fathom actor similarity in the mind of industry practitioners since actors who can replace each other must be considered similar in some aspects. After movie production has begun, casting decision may still change due to reasons such as injuries or artistic differences. For example, Nicole Kidman withdrew from the production of \emph{Panic Room} due to an injury and Jodie Foster took her role. Sequels and reboots may also use different actors for the same role. For example, Bruce Banner/Hulk was first played by Edward Norton (\textit{The Incredible Hulk}) and later by Mark Ruffalo (\textit{The Avengers}). 

We check known cases against the best cast prediction model (JGE-AGT). In the \textit{Dark Knight} trilogy, Maggie Gyllenhaal replaced Katie Holmes. Among more than two thousand actors, Gyllenhaal is the $55^{\text{th}}$ nearest neighbor (NN) of Holmes and Holmes is the $46^{\text{th}}$ NN of Gyllenhaal, making them the top 2.5\% most similar to each other. See Figure \ref{fig:qualeval} for more examples.

\section{Conclusions}

Little attention was paid to understanding actors in the narrative understanding literature, yet actors are correlated with the characters they play and may provide useful information for downstream applications such as content recommendation \cite{Yu2014personalized}. We present a joint representation learning algorithm for movies, actors, and personae using Gaussian embeddings that explicitly account for semantic uncertainty and actor versatility. The algorithm substantially surpasses TransE at cast list prediction by 331 in mean rank; it attains 59.72\% agreement with human judgment on actors' versatility in an unsupervised setting. To our knowledge, this is the first successful attempt at predicting actors' versatility. 

Despite errors in text processing and low coverage, automatically identified persona topics lead to consistent and statistically significant improvements on both cast prediction and versatility ranking. When compared with age and gender information, the results suggest the tasks offer complementary evaluation metrics for persona. Research on automatic persona identification is still in an early stage. By providing automatic evaluation methods, we believe this paper lays the groundwork for further advancement on narrative understanding.

%Representation learning extracts generic feature representations are extracted and may facilitate subsequent tasks. 
%The learned representation can inform downstream applications such as content recommendation and economic analysis of the motion picture industry \cite{Kumb2007}. 

%\section*{Acknowledgments}
%We gratefully acknowledge Stephan Mandt and Robert Bamler for valuable discussion. 

\bibliographystyle{aaai}
\bibliography{references}

\end{document}